# On the origin of topotactic reduction effect for superconductivity in infinite-layer nickelates

Shengwei Zeng[1, #, *], Chi Sin Tang[2,3, #, *], Zhaoyang Luo[2, #], Lin Er Chow[2], Zhi Shiuh Lim[1], Saurav Prakash[2], Ping Yang[3], Caozheng Diao[3], Xiaojiang Yu[3], Zhenxiang Xing[1], Rong Ji[1], Xinmao Yin[4], Changjian Li[5], X. Renshaw Wang[6], Qian He[7], Mark B. H. Breese[2,3], A. Ariando[2,*], Huajun Liu[1,*]

[1]Institute of Materials Research and Engineering (IMRE), Agency for Science, Technology and Research (A*STAR), 2 Fusionopolis Way, Innovis #08-03, Singapore 138634, Republic of Singapore.
[2]Department of Physics, Faculty of Science, National University of Singapore, Singapore 117551, Republic of Singapore.
[3]Singapore Synchrotron Light Source (SSLS), National University of Singapore, 5 Research Link, Singapore 117603, Republic of Singapore.
[4]Shanghai Key Laboratory of High Temperature Superconductors, Physics Department, Shanghai University, Shanghai 200444, China.
[5]Department of Materials Science and Engineering, Southern University of Science and Technology, Shenzhen 518055, Guangdong, China.
[6]Division of Physics and Applied Physics, School of Physical and Mathematical Sciences, Nanyang Technological University, Singapore 637371, Republic of Singapore.
[7]Department of Materials Science and Engineering, National University of Singapore, Singapore 117575, Republic of Singapore.

[#]The authors contributed equally to this work.
*To whom correspondence should be addressed.
E-mail: zeng_shengwei@imre.a-star.edu.sg; slscst@nus.edu.sg; ariando@nus.edu.sg; liu_huajun@imre.a-star.edu.sg

**Topotactic reduction utilizing metal hydrides as reagents emerges as an effective approach to achieve exceptionally low oxidization states of metal ions and unconventional coordination networks. This method opens avenues to the development of entirely new functional materials, with one notable example being the infinite-layer nickelate superconductors. However, the reduction effect on the atomic reconstruction and electronic structures – crucial for superconductivity – remains largely unresolved. We design two sets of control $Nd_{0.8}Sr_{0.2}NiO_2$ thin films and implement secondary ion mass spectroscopy to highlight the absence of reduction-induced hydrogen intercalation. X-ray absorption spectroscopy shows a significant linear dichroism with dominant Ni $3d_{x2-y2}$ orbitals on superconducting samples, indicating a Ni single-band nature of infinite-layer nickelates. Consistent with the superconducting $T_c$, the Ni $3d$ orbitals asymmetry manifests a dome-like reduction duration dependence. Our results unveil the critical role of reduction in modulating the Ni-$3d$ orbital polarization and its impact on the superconducting properties.**

The discovery of superconductivity in infinite-layer nickelate $(La/Pr/Nd)_{1-x}(Sr/Ca)_xNiO_2$ thin films has stimulated intensive interests owing to its scientific richness in understanding the mechanism of high-$T_c$ superconductivity[1-7]. Subsequent observations of pressure- and strain-induced $T_c$ enhancement in infinite-layer films[8-10], superconductivity in quintuple-layer nickelate $Nd_6Ni_5O_{12}$ (ref.[11]), and $T_c$ near 80 K in tri-layer compounds $La_3Ni_2O_7$ (ref.[12]) further suggest nickelates as an alternatively promising route towards high-$T_c$ superconductors. Even though it is resemblant to high-$T_c$ cuprates in $3d^9$ electron configuration and layered crystal structure, the infinite-layer nickelates showed different features such as Mott-Hubbard scenario in band structure[13-15], absence of long-range antiferromagnetic order[16-18], complexed paring symmetry [5, 19-22]. Even for samples with the same chemical composition, the experimental characterization of electronic structure such as the charge density wave is controversial[23-26]. Furthermore, bulk materials showed absence of superconductivity despite the presence of infinite-layer phase[27, 28]. Such disputable observations could be attributed to the challenging material synthesis and topotactic reduction process[29].

The infinite-layer nickelates are realized by the $CaH_2$-facilitated reduction from perovskite precursors, during which atomic reconstruction/element intercalation may occur and significantly modify the electronic structures[30]. A notable effect of reduction has been recently reported to be hydrogen (H) intercalation (H is from the reagent $CaH_2$), showing a critical role for superconductivity in $Nd_{0.8}Sr_{0.2}NiO_2H_x$, with zero resistivity only found in a very narrow H-concentration window[31, 32]. However, the theoretical calculation showed that the electron-phonon coupling involved with H doping is too weak to support the observed $T_c$ (ref.[33]). Moreover, $Nd_{1-x}Eu_xNiO_2$ thin films and the superconductivity have been achieved through in-situ reduction using metallic aluminium where the hydrogen source is absent[34]. Furthermore, defects such as Ruddlesden–Popper-type faults emerge during the deposition of perovskite precursors[2, 4, 29]. These defects with poor crystalline quality and boundary may cause moisture absorption, resulting in the controversial characterization of the reduction effect and its role for superconductivity. Therefore, it is desirable to clarify the reduction effect on atomic reconstruction and electronic structure, and the role leading to superconductivity in pure infinite-layer nickelates. Here we design two sets of nickelate samples, the first set is 5-nm pure infinite-layer $Nd_{0.8}Sr_{0.2}NiO_2$ film capped by 5-nm $SrTiO_3$ as a protection layer (**referred as STO/NSNO**) and the other set is 20-nm thicker film with mixed phases (**referred as tNSNO**). The tNSNO is chose as control samples to investigate the H intercalation as a result of mixed phases in the films. Time-of-flight secondary ion mass spectrometry (TOF–SIMS) show that the H intercalation in the pure infinite-layer phase is negligible. X-ray absorption spectroscopy (XAS) on the pure infinite-layer phases reveal a strong association between the onset of superconductivity and the precise 3$d$-orbital polarization. A dome-like reduction time dependence of superconductivity in infinite-layer nickelate is observed.

**Topotactic reduction effect on the crystal structure**

Figure 1a shows the high-angle annular darkfield scanning transmission electron microscopy (HAADF-STEM) image of the STO/NSNO at the optimally reduced duration of 80 mins. A clear infinite-layer structure is observed with no obvious defect throughout the layer over a wide range, confirming high crystalline quality. However, a defect area, which is widely distributed throughout the top area of nickelate layer, is clearly observed for sample at the over-reduced duration of 720 mins (Fig. 1b). In the defect area, a stripe feature is manifested with additional atoms between two Nd atom planes (Fig. 1b, left). These could be the interstitial Ni/Nd atoms which take the position of original oxygen in $NiO_2$ plane (Fig. 1b, middle and right). The formation of interstitial atoms could be related to the creation of oxygen vacancies in $NiO_2$ plane induced during the over-reduction process, and therefore, the neighbouring Nd or Ni atoms drift into the vacancy sites. The formation of such defects significantly influences the Ni 3$d$ orbital polarization and superconducting properties, which will be discussed in the following text. The reduction effect on the evolution of lattice structure can be seen from the change of $c$-axis lattice constants, $c$, which is calculated from X-ray diffraction (XRD) scans (Fig. 1c and Supplementary Fig. S1). With increasing reduction time, $c$ decreases and reach smallest value of ~3.331 Å, and then increase at longer reduction. The increase of $c$ at longer time of 360 and 720 mins is due to the presence of more oxygen vacancies and interstitial Ni/Nd. For tNSNO thin films, similar trend is seen with smallest $c$ at reduction time of 80 min and larger $c$ at longer reduction (Supplementary Fig. S1). In contrast to the infinite-layer phase in STO/NSNO samples, defects with Ruddlesden–Popper-type secondary phase is seen at the top section of the tNSNO sample[2, 4, 29] (Supplementary Fig. S1).

**Topotactic reduction effect on the electrical transport properties**

Figure 1d shows the resistivity *versus* temperature ($\rho$-$T$) curves of STO/NSNO thin films at different reduction time. The film reduced for 40 mins is highly insulating with the resistance out of the measurement limit. The resistivity decreases for the film reduced for 50 min but shows insulating behaviour throughout the temperature range below 300 K. By increasing reduction time to 80-360 mins, the samples become superconducting with the $\rho$-$T$ curves showing metallic behaviour at high temperature. Sharp superconducting transition and the highest critical temperature, $T_c$, are achieved at the reduction time of 80 and 180 mins (see also Fig. 4). By further increasing the reduction time to 720 mins, the sample shows metallic behaviour at high temperature and becomes weakly insulating behaviour at low temperature, which is similar the over-doped infinite-layer nickelates[2-4, 6, 7]. The reduction effect on superconducting property is summarized as a phase diagram in Fig. 4 (See also Supplementary Fig. S5), showing the change of $T_c$ versus reduction time. The phase diagram can be separated to under-reduced (insulating behaviour), optimally reduced (superconducting with highest $T_c$) and over-reduced (superconducting with lower $T_c$ and weekly insulating) regions. The reduction on tNSNO samples show a similar effect with superconducting transition for 80 mins and a weakly insulating behaviour for 720 mins (Fig. 1e). Figure 1f shows the temperature dependence of the normal-state Hall coefficients ($R_H$) of the STO/NSNO thin films. The $R_H$ shows a change from negative to positive signs at a low temperature for optimally reduced samples[1, 2, 4], while it remains negative for under- and over-reduced samples. The reduction duration dependence of the $R_H$ at 20 K and 300 K is plotted in Fig. 1g, clearly showing a sign change with increasing reduction time. These suggests a change of the multiband structures upon changing the reduction duration.

**H intercalation investigation by TOF-SIMS**

Figure 2a and 2b show the TOF-SIMS signals of H profile for bare STO substrate, STO/NSNO and tNSNO thin films at different reduction time. The Ni, Ti and Nd profiles of STO/NSNO and tNSNO at a reduction time of 80 mins are plotted in Fig. 2c and 2d, respectively, clearly indicating the well-defined interfaces and layer structures between nickelate and STO. The three-dimensional (3D) element mappings show that the elements of the nickelate film are uniformly distributed with no chemical segregation within the measurement size of 100 μm

x 100 μm (Fig. 2e). All other SIMS data are shown in Supplementary Fig. S2 and Fig. S3. With STO capping, the as-grown and reduced STO/NSNO samples at different durations show comparable H signal intensity with exception for the surface and interface. This indicates that the effect of $CaH_2$ reduction on H intercalation in infinite-layer film is negligible. Moreover, the difference in H signal intensity between the nickelate films and the background of STO substrates is not obvious. This is in contrast to previous observation in the reference[31], where the H signal intensity in single-layer $Nd_{0.8}Sr_{0.2}NiO_2$ films is two orders of magnitude higher than that in STO. For comparison, we also performed TOF-SIMS measurements on $Nd_{0.8}Sr_{0.2}NiO_2$ film without STO capping, i.e., 20-nm tNSNO (Fig. 2b). For these samples, the pure infinite-layer structure with thickness up to ~10 nm could be achieved at the bottom area of the film near STO substrate, whereas the top area above ~10 nm consists of mixed phases[2, 4, 29] (Supplementary Fig. S1). One can see that the H signal intensity is higher on the surface, and gradually decreases into the film. At the bottom of the film (sputtering time from 200 to 400 sec), the intensity is comparable to that of the STO/NSNO film and STO substrates. Overall, the low H intensity in STO/NSNO and bottom area of tNSNO reduced over a wide range of durations indicate that the reduction-induced H intercalation in pure infinite-layer phase is negligible. Therefore, this highlights that H doping is not necessary for superconductivity in infinite-layer nickelates. In contrast, higher H intensity at the top area of tNSNO, possibly indicate the moisture absorption in the film due to the presence of defects.

It is worth noting that reduction-induced H intercalation was reported to be in $NdNiO_xH_y$ film with defect-fluorite structure, while H is absent in infinite-layer phase[35]. Moreover, it has also been reported that no hydrogen in $LaNiO_2$ polycrystalline powder is detected using neutron diffraction, and the extra hydrogen might be confined to polycrystalline grain boundaries or secondary-phase precipitate[36]. Metal hydride reduction induced H intercalation was reported in other perovskites such as SrVO$_x$ (ref.[30, 37]). However, the reduction temperature (~ 600 °C) and duration (~ 24 hours) required for topotactic transition in SrVO$_x$ is much higher and longer, compared to those for nickelates (240-340 °C and several hours)[30]. Therefore, to avoid H intercalation in infinite-layer nickelate, precursor with pure perovskite phase and lower reduction temperature is desirable.

**X-ray absorption spectroscopy**

Having deduced that the H intercalation is negligible, we then investigate the effects of reduction on the nickelate electronic structures. The reduction duration is found to play a directly relevant role to the superconductivity of infinite-layer nickelate. To carefully characterize the intrinsic electronic structures for pure infinite-layer phase[2, 4, 29], the STO/NSNO samples were selected for XAS measurements. Figure 3a shows an example of Ni $L_{2,3}$ edge XAS of the sample reduced for 180 mins where the spectra are taken at incident angles of 20° ($\varepsilon \perp a$, out-of-plane, $\varepsilon$ indicates electric field and $a$ indicates the NiO$_2$ plane) and 90° ($\varepsilon /\!/ a$, in-plane). The dominant absorption peaks at ~853 eV and ~870 eV are clearly observed, corresponding to the $2p^6_{3/2}3d^9 \Rightarrow 2p^5_{3/2}3d^{10}$ and $2p^6_{1/2}3d^9 \Rightarrow 2p^5_{1/2}3d^{10}$ transitions, respectively. The additional shoulder on the higher energy side relative to the main peaks show significantly lower intensity as compared to that in previous reports[38]. The light polarization effect shows a significant linear dichroism with a larger absorption as $\varepsilon$ is aligned in-plane, as compared to that when $\varepsilon$ is aligned out-of-plane. The Ni $L_{2,3}$ edges XAS and linear dichroism for other STO/NSNO samples reduced at different durations are plotted together in Supplementary Fig. S4. The spectra linear dichroism, $\Delta I = I(\varepsilon /\!/ a) - I(\varepsilon \perp a)$, are shown in Fig. 3b where the Ni-$L_3$ peak dichroism are plotted with respect to the reduction duration in the inset of Fig. 3b. For the sample reduced for 40 mins, there is a slight linear dichroism. As the reduction duration increases, there is a corresponding increase in $\Delta I$ and it maximizes for the sample that has undergone a reduction duration of 180 mins. Beyond which, the $\Delta I$ decreases as the samples become over-reduced when the reduction process lasted for 360 and 720 mins, respectively. To further characterize the linear dichroism, the Ni-$L_3$ peak asymmetry percentage, $\Delta I\% = [I(\varepsilon /\!/ a) - I(\varepsilon \perp a)] / [I(\varepsilon /\!/ a) + I(\varepsilon \perp a)]$, is plotted as a function of reduction duration in Fig. 4. The asymmetry shows a dome-shape behaviour with respect to reduction time, with the largest asymmetry of ~ 69 % at 180 mins, which is consistent with the change of $T_c$.

**Discussion**

The clear correlation between the XAS linear dichroism and reduction duration is a strong indication that the reduction process has a direct impact on the Ni 3$d$ orbital polarization, and this in turn is critical to the superconducting properties of the infinite-layer nickelates. According to the crystal field configuration, the energy level of Ni 3$d_{z2}$ orbital is lowered to be comparable to that of the $t_{2g}$ orbitals under NiO$_2$ square planar coordination, and therefore, 3$d_{x2-y2}$ orbital becomes isolated from other 3$d$ orbitals[39]. The optimally reduced condition enables the realization of this perfect square planar field, leading to a single Ni 3$d_{x2-y2}$ band with two-dimensional nature which favours the occurrence of superconductivity[39, 40]. This is consistent with the smallest $c$-axis lattice constant and the largest orbital asymmetry at the optimally reduced region (Fig. 4 and Supplementary Fig. S5). In the under-reduced condition, with the incomplete transition to the infinite-layer structure, residual apical oxygens with a shorter Ni-O apical distance are present[41]. This results in the lifting of 3$d_{z2}$ level and the resultant admixing of the $3d_{x2-y2}$ and $3d_{z2}$ orbitals. The doped holes and unoccupied states reside in this mixed band and thus, orbital asymmetry is small. In the over-reduced condition, the creation of oxygen vacancies in NiO$_2$ plane and the resultant formation of interstitial Ni/Nd atoms lead to deviation of the Ni-coordination from the long range-order planar square configuration. Likewise, this structural deviation results in the hybridization of the $3d_{x2-y2}$ and $3d_{z2}$ orbitals. Hence, in both the under- and over-reduced conditions, the presence of residual apical oxygen and interstitial atoms in NiO$_2$ planes have detrimental effects on the realization of superconductivity due to the hybridization effects of both the $3d_{x2-y2}$ and $3d_{z2}$ orbitals. The weakening of single-band feature due to residual apical oxygen and its detrimental effect to superconductivity in infinite-layer nickelate have also been demonstrated at the NdNiO$_2$/STO interfaces[39, 42-44].

The reduction effect on the orbital polarization is consistent with the observation that multiband structures become more pronounced in the under- and over-reduced regimes. As shown in Fig. 1f and 1g, $R_H$ remains negative in both the under- and over-reduced regimes while there is a change in sign at optimally reduced regime. It should be noted that the multiband structures may originate from the rare-earth element $5d_{xy}$ and $5d_{z2}$ orbitals, and the reduction effect on these orbitals is worthy of further investigation[39, 45]. With respect to infinite-layer nickelate bulks, superconductivity is absent even though that the perfect diamagnetism and high critical current density measured on thin film indicate the bulk nature of superconductivity in infinite-layer nickelates[1, 46]. In bulk materials, the reduction duration required to obtain the infinite-layer structure is more than 20 hours[27, 28, 47]. Such prolonged reduction duration makes it vulnerable to the formation of in-plane oxygen vacancies and interstitial atoms, which may be the reason for absence of superconductivity in the bulk form[27, 28, 47]. In summary, we highlight that the CaH$_2$ reduction-induced H intercalation is negligible in pure infinite-layer phase, thus not critical to superconductivity. The reduction significantly influences the Ni 3$d$ orbital polarization with dominant 3$d_{x^2-y^2}$ orbital in optimally reduced condition, which is critical for realization of superconductivity in infinite-layer nickelates.

**Methods**

The nickelate thin film preparation is the same as described in our previous reports[4, 7]. Two sets of samples were grown on TiO$_2$-terminated (001) SrTiO$_3$ (STO) substrates by a pulsed laser deposition (PLD) system using a 248-nm KrF excimer laser. The first set is an ultrathin 5.7-nm perovskite Nd$_{0.8}$Sr$_{0.2}$NiO$_3$ thin film (after reduction the thickness of infinite-layer Nd$_{0.8}$Sr$_{0.2}$NiO$_2$ is ~5 nm) capped with 5-nm STO top layer (**referred as STO/NSNO**); the other set is a 20-nm thick Nd$_{0.8}$Sr$_{0.2}$NiO$_3$ thin film without capping layer (**referred as tNSNO**). The deposition temperature and oxygen partial pressure $P_{O2}$ for both Nd$_{0.8}$Sr$_{0.2}$NiO$_3$ and STO are 600 °C and 150 mTorr. The laser energy density on the target surface was set to be 2 Jcm$^{-2}$. After deposition, the samples were cooled down to room temperature at 150 mTorr and a rate of 8 °C/min. For each set of samples, the nickelate film was deposited on the STO substrate with a size of 10 x 10 mm$^2$, and then cut into small pieces with a size of around 2.5 x 1.5 mm$^2$ for reduction. The reduction was done in PLD chamber at 340 °C for different duration of 40, 50, 80, 180, 360 and 720 min with a heating rate and cooling rate of 25 °C /min. During the reduction, the samples were embedded with about 0.2 g of CaH$_2$ powder and wrapped in aluminium foil.

The electrical transport measurement was performed using a Quantum Design Physical Property Measurement System. The connection for the electrical transport measurement was made by Al ultrasonic wire bonding. The X-ray diffraction (XRD) measurement was done in the X-ray Diffraction and Development (XDD) beamline at Singapore Synchrotron Light Source (SSLS) with an X-ray wavelength of $\lambda$ = 1.5404 Å. The X-ray absorption spectroscopy (XAS) measurements were performed using linearly polarized X-ray from the Soft X-Ray—Ultraviolet (SUV) beamline and the Surface, Interface, and Nanostructure Science (SINS) beamline at SSLS. A total electron yield (TEY) detection method is used during the measurements. The incidence angle of X-rays refers to the sample surface was varied by rotating the polar angle of the sample. The spectra were measured at the incident angle of 20º ($\varepsilon \perp a$) and 90º ($\varepsilon /\!/ a$). The spectra were normalized to the integrated intensity at the tail of the spectra after subtracting an energy-independent background.

The high-angle annular dark-field scanning transmission electron microscopy (HAADF-STEM) imaging was carried out at 200 kV using a JEOL ARM200F microscope and the cross-sectional TEM specimens were prepared by a focused ion beam machine (FEI Versa 3D).

The element depth profiles were measured using time-of-flight secondary ion mass spectrometry (TOF–SIMS). The measurement tool is TOF-SIMS-5 instrument (IONTOF GmbH). The measurements were done in dual beam mode by two ion beams with $Cs^+$ ions as sputter beam (500 μm x 500 μm) and $Bi^+$ ions as analysis beam (100 μm x 100 μm). The analysis beam with a smaller size in the centre of the sputter beam is chose to avoid disturbance from the edge. The signals were collected by negative secondary ion.

**Acknowledgments**

S.W.Z., Z.S.L., and H.J.L. acknowledge the RIE2025 MTC Individual Research Grants (M22K2c0084), National Research Foundation Competitive Research Program (NRF-CRP28-2022-0002), Career Development Fund (C210812020) and Central Research Fund from the Agency for Science, Technology and Research (A*STAR) for the funding support. The work at NUS and the authors are supported by the Ministry of Education (MOE), Singapore, under its Tier-2 Academic Research Fund (AcRF) (Grants No. MOE-T2EP50121-0018 and MOE T2EP50123-0013) and the National Research Foundation (NRF) of Singapore under its NRF-ISF joint program (Grant no. NRF2020-NRF-ISF004-3518). X.R.W. acknowledges support from Singapore Ministry of Education under its Academic Research Fund (AcRF) Tier 2 (Grant No. MOE-T2EP50120-0006). X.M.Y. acknowledges financial support by National Natural Science Foundation of China (Grant No.12374378). C.S.T acknowledges the support from the NUS Emerging Scientist Fellowship. The authors would like to acknowledge the Singapore Synchrotron Light Source for providing the facility necessary for conducting the research. The Laboratory is a National Research Infrastructure under the National Research Foundation, Singapore. Any opinions, findings, and conclusions or recommendations expressed in this material are those of the author(s) and do not reflect the views of National Research Foundation, Singapore.

**Author contributions**

S.W.Z., H.J.L., A.A., C.S.T. and Z.Y.L. conceived the main idea. S.W.Z., Z.Y.L., L.E.C., Z.S.L. and S.P. prepared the thin films and conducted the electrical measurements. C.S.T., S.W.Z. and P.Y. conducted the XRD measurements. C.S.T., C.Z.D., X. J. Y. and M.B.H.B. conducted the XAS measurements. Z.Y.L., Q. H. and C.J.L. conducted the STEM measurements. Z.X.X. and R.J. conducted the TOF-SIMS measurements. X.M.Y., C.J.L. and X.R.W. provide insight to manuscript. S.W.Z., H.J.L., A.A., C.S.T. and Z.Y.L. wrote the manuscript with contributions from all authors. All authors have discussed the results and the interpretations.

## Competing interests

The authors declare no competing interests.

## References

[1] Li, D. et al. Superconductivity in an infinite-layer nickelate. *Nature* **572**, 624-627 (2019).
[2] Li, D. et al. Superconducting dome in $Nd_{1-x}Sr_xNiO_2$ infinite layer films. *Phys. Rev. Lett.* **125**, 027001 (2020).
[3] Osada, M. et al. Phase diagram of infinite layer praseodymium nickelate $Pr_{1-x}Sr_xNiO_2$ thin films. *Phys. Rev. Mater.* **4**, 121801 (2020).
[4] Zeng, S. et al. Phase diagram and superconducting dome of infinite-layer $Nd_{1-x}Sr_xNiO_2$ thin films. *Phys. Rev. Lett.* **125**, 147003 (2020).
[5] Gu, Q. et al. Single particle tunneling spectrum of superconducting $Nd_{1-x}Sr_xNiO_2$ thin films, *Nat. Commun.* **11**, 6027 (2020).
[6] Osada, M. Nickelate superconductivity without rare‐earth magnetism: (La, Sr) $NiO_2$. *Adv. Mater.* **33**, 2104083 (2021).
[7] Zeng, S. et al. Superconductivity in infinite-layer nickelate $La_{1-x}Ca_xNiO_2$ thin films. *Sci. Adv.* **8**, eabl9927 (2022).
[8] Wang, N. et al. Pressure-induced monotonic enhancement of $T_c$ to over 30 K in superconducting $Pr_{0.82}Sr_{0.18}NiO_2$ thin films. *Nat. Commun.* **13**, 4367 (2022).
[9] Lee, K. et al. Linear-in-temperature resistivity for optimally superconducting (Nd, Sr) $NiO_2$. *Nature* **619**, 288-292 (2023).
[10] Ren, X. et al. Possible strain-induced enhancement of the superconducting onset transition temperature in infinite-layer nickelates. *Commun.Phys*. **6**, 341 (2023).
[11] Pan G. A. et al. Superconductivity in a quintuple-layer square-planar nickelate. *Nat. Mater.* **21**, 160-164 (2022).
[12] Sun, H. et al. Signatures of superconductivity near 80 K in a nickelate under high pressure. *Nature*, **621**, 493-498 (2023).
[13] Hepting, M. et al. Electronic structure of the parent compound of superconducting infinite-layer nickelates. *Nat. Mater.* **19**, 381-385 (2020).

[14] Chen, Z. et al. Electronic structure of superconducting nickelates probed by resonant photoemission spectroscopy. *Matter*, **5**, 1806-1815 (2022).
[15] Jiang, M., Berciu, M., Sawatzky, G. A., Critical Nature of the Ni Spin State in Doped $NdNiO_2$. *Phys. Rev. Lett.* **124**, 207004 (2020).
[16] Lu, H. et al. Magnetic excitations in infinite-layer nickelates. *Science*, **373**, 213-216 (2021).
[17] Zhou, X. et al. Antiferromagnetism in Ni‑Based Superconductors. *Adv. Mater.* **34**, 2106117 (2022).
[18] Fowlie, J. et al. Intrinsic magnetism in superconducting infinite-layer nickelates. *Nat. Phys.* **18**, 1043-1047 (2022).
[19] Chow, L. E. et al. Pairing symmetry in infinite-layer nickelate superconductor. Preprint at https://arxiv.org/abs/2201.10038 (2022).
[20] Harvey, S. P. et al. Evidence for nodal superconductivity in infinite-layer nickelates. Preprint at https://arxiv.org/abs/2201.12971 (2022).
[21] Cheng, B. et al. Evidence for d-wave superconductivity of infinite-layer nickelates from low-energy electrodynamics. *Nat. Mater.* https://doi.org/10.1038/s41563-023-01766-z (2024).
[22] Wang, Z., Zhang, G.-M., Yang, Y.-F., Zhang, F.-C. Distinct pairing symmetries of superconductivity in infinite-layer nickelates. *Phys. Rev. B*, **102**, 220501 (2020).
[23] Tam, C. C. et al. Charge density waves in infinite-layer $NdNiO_2$ nickelates. *Nat. Mater.* **21**, 1116-1120 (2022).
[24] Krieger, G. et al. Charge and spin order dichotomy in $NdNiO_2$ driven by the capping layer. *Phys. Rev. Lett.* **129**, 027002 (2022).
[25] Rossi, M. et al. A broken translational symmetry state in an infinite-layer nickelate. *Nat. Phys.* **18**, 869-873 (2022).
[26] Parzyck, C. et al. Absence of 3a0 charge density wave order in the infinite-layer nickelate $NdNiO_2$. *Nat. Mater.* https://doi.org/10.1038/s41563-024-01797-0 (2024).
[27] Li, Q. et al. Absence of superconductivity in bulk $Nd_{1-x}Sr_xNiO_2$. *Commun. Mater.* **1**, 16 (2020).
[28] Wang, B.-X. et al. Synthesis and characterization of bulk $Nd_{1-x}Sr_xNiO_2$ and $Nd_{1-x}Sr_xNiO_3$. *Phys. Rev. Mater.* **4**, 084409 (2020).
[29] Lee, K. B. et al. Aspects of the synthesis of thin film superconducting infinite-layer nickelates. *APL Mater.* **8**, 041107 (2020).
[30] Meng, Z. et al. Topotactic Transition: A Promising Opportunity for Creating New Oxides. *Adv. Funct. Mater.* **33**, 2305225 (2023).
[31] Ding, X. et al. Critical role of hydrogen for superconductivity in nickelates. *Nature*, **615**, 50-55 (2023).
[32] Qin, C., Jiang, M., Si, L., Effects of different concentrations of topotactic hydrogen impurities on the electronic structure of nickelate superconductors. *Phys. Rev. B* **108**, 155147 (2023).
[33] Di Cataldo, S., Worm, P., Si, L., Held, K. Absence of electron-phonon-mediated superconductivity in hydrogen-intercalated nickelates. Preprint at https://arxiv.org/abs/2304.03599 (2023).
[34] Wei, W., Vu, D., Zhang, Z., Walker, F. J., Ahn, C. H. Superconducting $Nd_{1-x}Eu_xNiO_2$ thin films using in situ synthesis. *Sci. Adv.* **9**, eadh3327 (2023).
[35] Onozuka, T., Chikamatsu, A., Katayama, T., Fukumura, T., Hasegawa, T. Formation of defect-fluorite structured $NdNiO_xH_y$ epitaxial thin films via a soft chemical route from $NdNiO_3$ precursors. *Dalton Trans.* **45**, 12114-12118 (2016).
[36] Puphal, P. et al. Investigation of hydrogen incorporations in bulk infinite-layer nickelates. *Front. Phys.* **10**, 115 (2022).
[37] Katayama, T. et al. Epitaxial growth and electronic structure of oxyhydride $SrVO_2H$ thin films. *J. Appl. Phys.*, **120**, 085305 (2016).
[38] Rossi, M. et al. Orbital and spin character of doped carriers in infinite-layer nickelates. Phys. Rev. B, 104, L220505 (2021).
[39] Nomura, Y., Arita, R. Superconductivity in infinite-layer nickelates. *Rep. Prog. Phys*. **85**, 052501 (2022).

[40] Kitatani, M. et al. Nickelate superconductors—a renaissance of the one-band Hubbard model. *npj Quantum Mater.* **5**, 59 (2020).
[41] Han, K. et al. Metal–insulator–superconductor transition in nickelate-based heterostructures driven by topotactic reduction. *Appl. Phys. Lett.* **123**, 182601 (2023).
[42] He, R. et al. Polarity-induced electronic and atomic reconstruction at $NdNiO_2$/$SrTiO_3$ interfaces. *Phys. Rev. B*, **102**, 035118 (2020).
[43] Bernardini, F., Cano, A. Stability and electronic properties of $LaNiO_2$/$SrTiO_3$ heterostructures. *J. Phys. Mater.* **3**, 03LT01 (2020).
[44] Goodge, B. H. et al. Resolving the polar interface of infinite-layer nickelate thin films. *Nat. Mater.* **22**, 466-473 (2023).
[45] Botana, A. S., Norman, M. R., Similarities and Differences between $LaNiO_2$ and $CaCuO_2$ and Implications for Superconductivity. *Phys. Rev. X* **10**, 011024 (2020).
[46] Zeng, S. et al. Observation of perfect diamagnetism and interfacial effect on the electronic structures in infinite layer $Nd_{0.8}Sr_{0.2}NiO_2$ superconductors. *Nat. Commun.* **13**, 743 (2022).
[47] Puphal, P. et al. Topotactic transformation of single crystals: From perovskite to infinite-layer nickelates. *Sci. Adv.* **7**, eabl8091 (2021).

# Figures and Captions

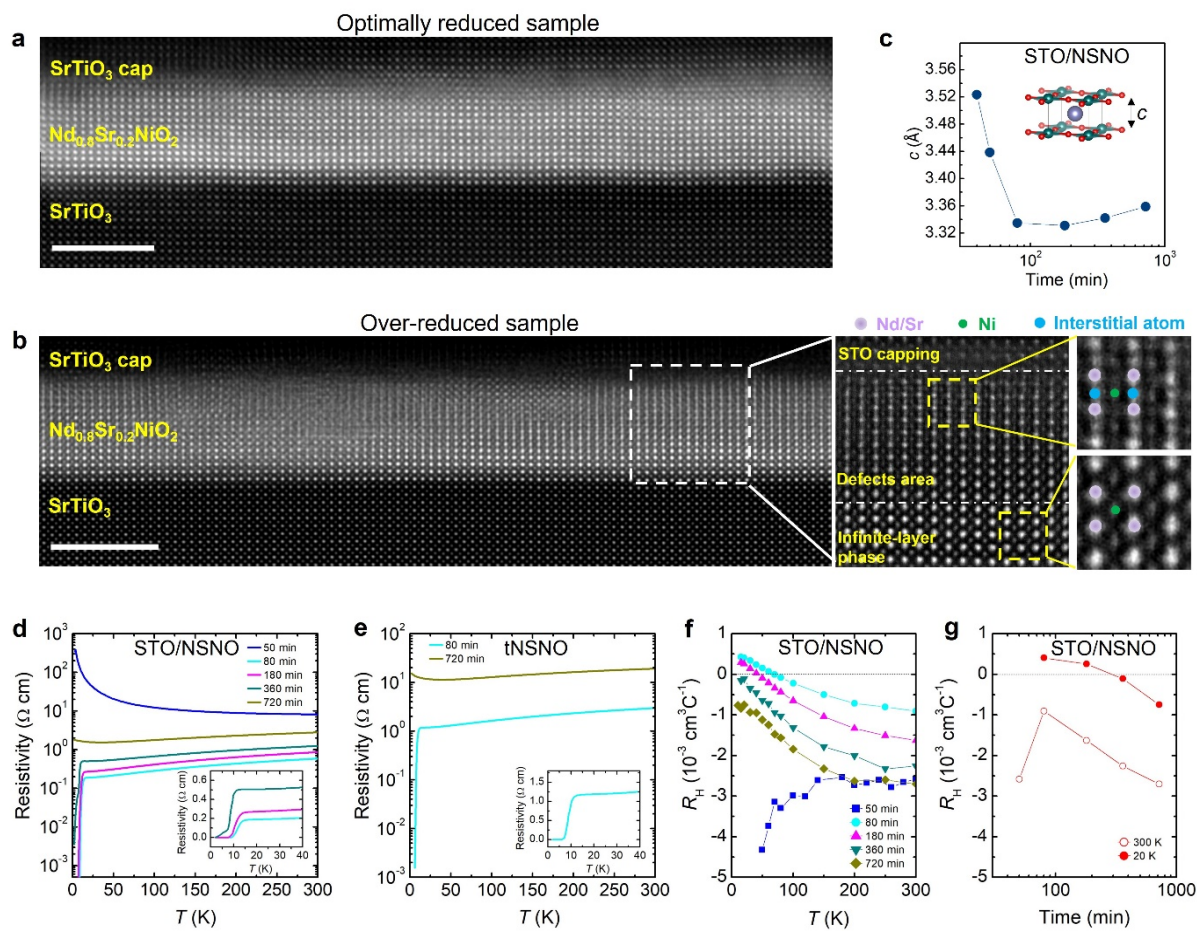

**Fig. 1 Topotactic reduction effect on the crystalline structure and electrical transport property of infinite-layer $Nd_{0.8}Sr_{0.2}NiO_2$ thin films.** (**a**) HAADF-STEM cross-sectional image of the 5-nm infinite-layer $Nd_{0.8}Sr_{0.2}NiO_2$ capped by a $SrTiO_3$ layer (referred as STO/NSNO) at the optimally reduced duration of 80 mins. A clear infinite-layer structure is observed throughout the layer over a wide range. (**b**) HAADF-STEM cross-sectional image of the STO/NSNO sample at the over-reduced duration of 720 mins (left), magnified view of the region marked by the white dashed box (middle) and regions marked by the yellow dashed box (right). The Ni atoms are shaded in green, Nd/Sr in lavender and the interstitial Ni/Nd in light blue. A defect area is observed in the magnified image with possible interstitial Ni/Nd at the oxygen vacancy sites in $NiO_2$ plane created during the over-reduction process. This defect area is widely distributed throughout the nickelate film. (**c**) The room-temperature $c$-axis lattice constants, $c$, calculated from XRD θ-2θ scans (Supplementary Fig. S1), plotted as a function of reduction time for the STO/NSNO samples. The inset is the crystal structure of infinite-layer nickelate. (**d**) The

resistivity *versus* temperature ($\rho$-$T$) curves of STO/NSNO thin films in logarithmic scale for different reduction time of 50, 80, 180, 360 and 720 mins. (**e**) The $\rho$-$T$ curves of the 20-nm Nd$_{0.8}$Sr$_{0.2}$NiO$_2$ thin films in logarithmic scale for different reduction time of 80 and 720 mins (referred as tNSNO). The insets of (**d**) and (**e**) show the zoomed-in and linear-scale $\rho$-$T$ curves of the superconducting samples. For both sets of samples, superconducting transition is seen at optimally reduction conditions of 80-360 mins while insulating behaviour is found at under- and over-reduced conditions. (**f**) The temperature dependence of the normal-state Hall coefficients $R_H$ of the STO/NSNO samples. (**g**) The $R_H$ of the STO/NSNO samples at 300 and 20 K as a function of reduction duration. Scale bars in (**a**) and (**b**) are 5 nm.

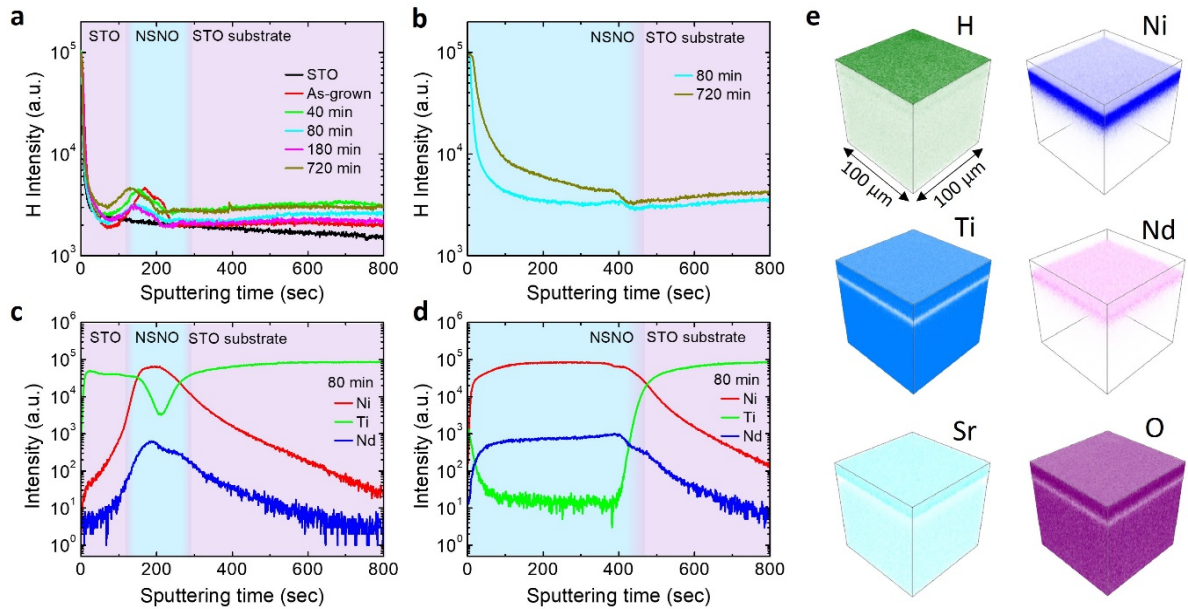

**Fig. 2 TOF-SIMS element profiles and 3-dimentional (3D) mapping.** (**a**) SIMS signals of H profile for STO, as-grown STO/NSNO and reduced STO/NSNO thin films at different reduction durations of 40, 80, 180, and 720 mins. The as-grown and reduced STO/NSNO samples show comparable H signal intensity, indicating negligible H intercalation during $CaH_2$ reduction process. (**b**) SIMS signals of H profile for tNSNO. Higher H signal intensity is visible at the top area of nickelate film with mixed phases. (**c**) Signals of Ni, Ti and Nd profile for STO/NSNO at reduction time of 80 min. (**d**) Signals of Ni, Ti and Nd profile for tNSNO at reduction time of 80 mins. Ni, Ti and Nd profile clearly indicate the well-defined interfaces and layer structures between nickelate and STO. (**e**) 3D mappings of H, Ni, Ti, Nd, Sr and O elements for the STO/NSNO reduced for 80 mins. Elements of the nickelate film are uniformly distributed with no chemical segregation. The elements H, Ni, Ti, Nd, Sr, O profiles are collected from the negative ions $H^-$, $Ni^-$, $TiO^-$, $NdO^-$, $SrO^-$, $O^{2-}$, respectively.

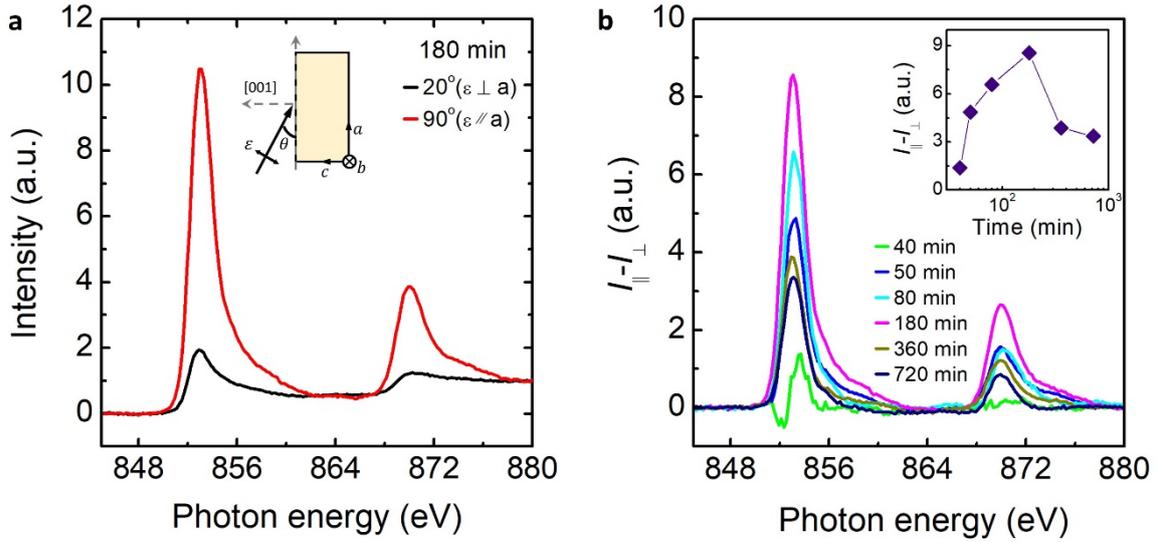

**Fig. 3 Topotactic reduction effect on X-ray absorption spectroscopy.** (**a**) Ni $L_{2,3}$ edge XAS of the STO/NSNO at reduction time of 180 mins. The spectra were collected at incident angle of 20° ($\varepsilon \perp a$, light polarization $\varepsilon$ vector is mostly perpendicular to the sample surface) and 90° ($\varepsilon \parallel a$, light polarization $\varepsilon$ vector is parallel to the sample surface). A significant x-ray linear dichroism (XLD) is observed with a larger absorption as $\varepsilon$ is aligned in-plane. The XAS of all other samples are shown in Supplementary Fig. S4. The inset shows the direction of photon polarization $\varepsilon$ and incident angle. (**b**) The intensity difference (XLD, $\Delta I = I(\varepsilon \parallel a) - I(\varepsilon \perp a)$) for STO/NSNO samples reduced at different time of 40, 50, 80, 180, 360 and 720 mins. The inset is Ni-$L_3$ peak intensity difference versus reduction time.

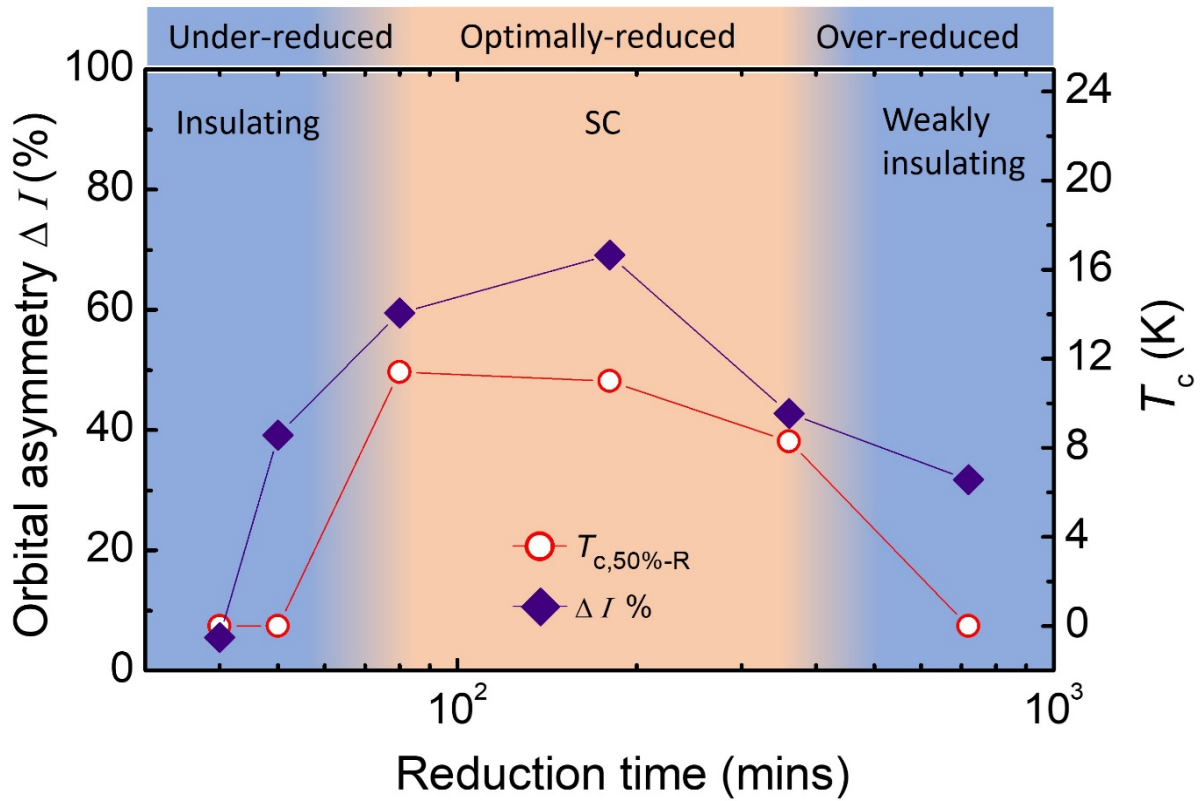

**Fig. 4 Reduction phase diagram of infinite-layer Nd$_{0.8}$Sr$_{0.2}$NiO$_2$ thin films.** The XAS spectral asymmetry at Ni $L_3$ peak, $\Delta I\% = [I(\varepsilon /\!/ a) - I(\varepsilon \perp a)] / [I(\varepsilon /\!/ a) + I(\varepsilon \perp a)]$, and the superconducting critical temperature $T_c$ versus reduction duration for STO/NSNO samples. The $T_{c\ 50\%\text{-}R}$ is defined to be the temperature at which the resistivity drops to 50 % of the value at 20 K (the onset of the superconductivity). The reduction time is in logarithmic scale. A consistent dome-shape behaviour with respect to the reduction time is seen for both $T_c$ and orbital asymmetry.

# Supplementary Information

# On the origin of topotactic reduction effect for superconductivity in infinite-layer nickelates

Shengwei Zeng[1, #,*], Chi Sin Tang[2,3, #,*], Zhaoyang Luo[2, #], Lin Er Chow[2], Zhi Shiuh Lim[1], Saurav Prakash[2], Ping Yang[3], Caozheng Diao[3], Xiaojiang Yu[3], Zhenxiang Xing[1], Rong Ji[1], Xinmao Yin[4], Changjian Li[5], X. Renshaw Wang[6], Qian He[7], Mark B. H. Breese[2,3], A. Ariando[2,*], Huajun Liu[1,*]

[1]Institute of Materials Research and Engineering (IMRE), Agency for Science, Technology and Research (A*STAR), 2 Fusionopolis Way, Innovis #08-03, Singapore 138634, Republic of Singapore.
[2]Department of Physics, Faculty of Science, National University of Singapore, Singapore 117551, Republic of Singapore.
[3]Singapore Synchrotron Light Source (SSLS), National University of Singapore, 5 Research Link, Singapore 117603, Republic of Singapore.
[4]Shanghai Key Laboratory of High Temperature Superconductors, Physics Department, Shanghai University, Shanghai 200444, China.
[5]Department of Materials Science and Engineering, Southern University of Science and Technology, Shenzhen 518055, Guangdong, China.
[6]Division of Physics and Applied Physics, School of Physical and Mathematical Sciences, Nanyang Technological University, Singapore 637371, Republic of Singapore.
[7]Department of Materials Science and Engineering, National University of Singapore, Singapore 117575, Republic of Singapore.

[#]The authors contributed equally to this work.
*To whom correspondence should be addressed.
E-mail: zeng_shengwei@imre.a-star.edu.sg; slscst@nus.edu.sg; ariando@nus.edu.sg; liu_huajun@imre.a-star.edu.sg

Fig. S1 Reduction effect on the crystalline structure of infinite-layer $Nd_{0.8}Sr_{0.2}NdO_2$ thin films.

Fig. S2 TOF-SIMS element profiles for 5-nm $Nd_{0.8}Sr_{0.2}NdO_2$ with 5-nm STO capping.

Fig. S3 TOF-SIMS element profiles and 3-dimentional (3D) mapping for 20-nm $Nd_{0.8}Sr_{0.2}NdO_2$.

Fig. S4 Reduction dependence of the X-ray absorption spectroscopy.

Fig. S5 Schematic of lattice structures and the reduction phase diagram of infinite-layer $Nd_{0.8}Sr_{0.2}NiO_2$ thin films.

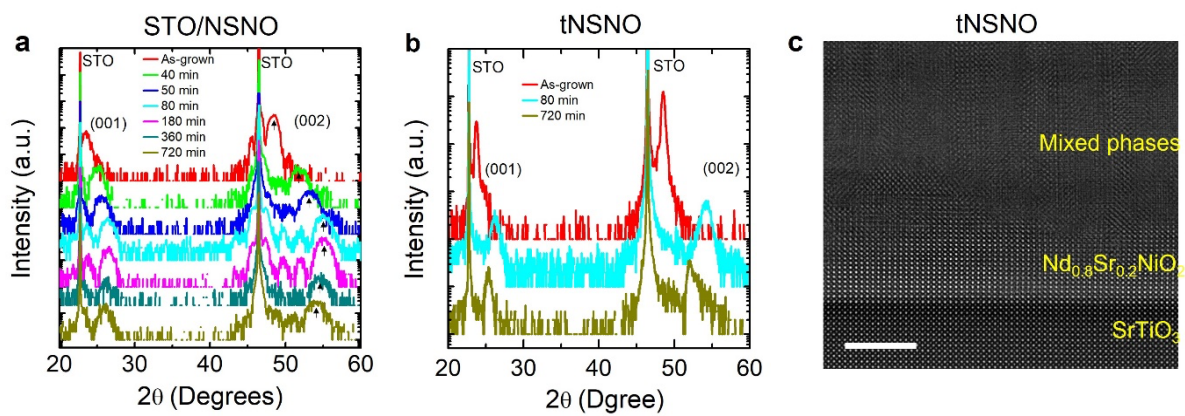

**Supplementary Fig. S1 Reduction effect on the crystalline structure of infinite-layer $Nd_{0.8}Sr_{0.2}NdO_2$ thin films.** (**a**) The XRD θ-2θ scans of the STO/NSNO samples at different reduction durations of 40, 50, 80, 180, 360 and 720 mins. The intensity is vertically offset for clarity. The black arrows serve as visual guides to indicate the changes in the diffraction peak positions. The clear transition of the diffraction peaks confirms the transformation from the perovskite to the infinite-layer structure after reduction. The Laue oscillations in the vicinity of the (002) peak indicate the high crystalline quality of the infinite-layer films. The shift of diffraction peak positions indicates that with increasing reduction time, the *c*-axis lattice constants, *c,* decreases and reach smallest value, and then increase at longer reduction (see also main text Fig. 1c). (**b**) The XRD θ-2θ scans of as-grown tNSNO and reduced tNSNO thin films at different reduction time of 80 and 720 mins. The *c* of tNSNO show a similar trend as that of STO/NSNO, with smallest value at reduction time of 80 mins and larger *c* at longer reduction of 720 mins. (**c**) HAADF-STEM cross-sectional image of the tNSNO thin film reduced for 80 mins. The infinite-layer structure is seen at the bottom section of the film near the STO substrate. At the top section, mixed phase with the Ruddlesden–Popper-type secondary phase is observed. Scale bar, 5 nm.

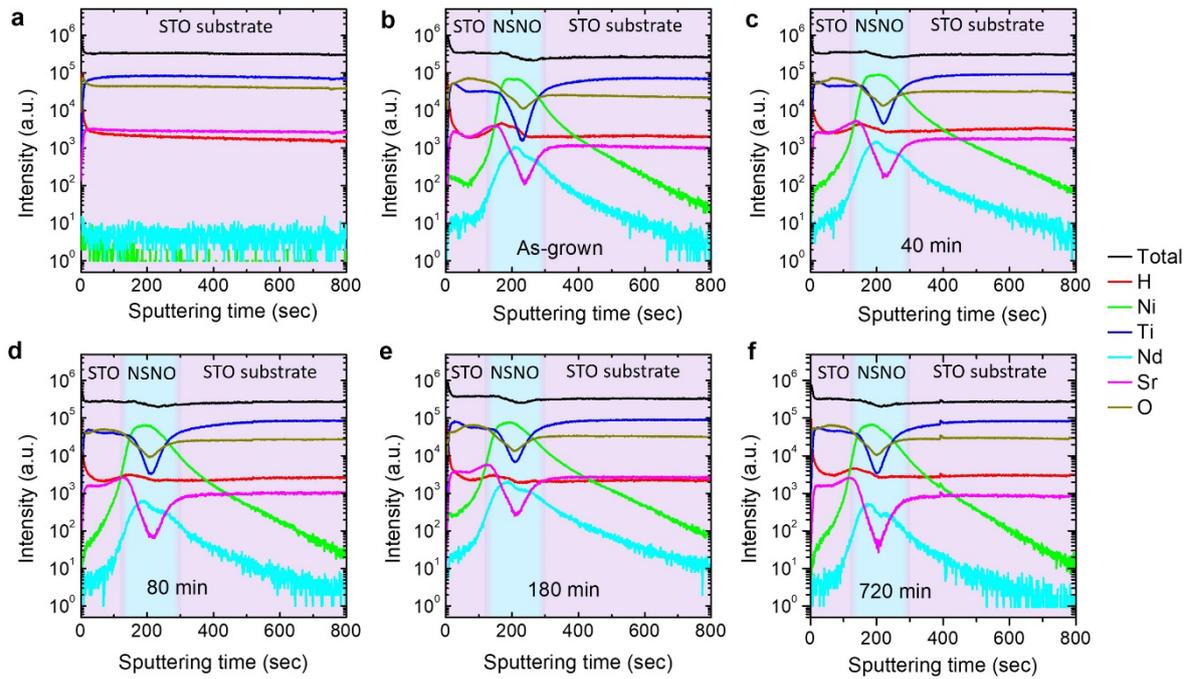

**Supplementary Fig. S2 TOF-SIMS element profiles for 5-nm Nd$_{0.8}$Sr$_{0.2}$NdO$_2$ with 5-nm STO capping.** SIMS signals of total, H, Ni, Ti, Nd, Sr and O profile for (**a**) STO, (**b**) as-grown STO/NSNO, and reduced STO/NSNO at (**c**) 40 min, (**d**) 80 min, (**e**) 180 min and (**f**) 720 min. The element profiles clearly indicate the well-defined interfaces and layer structures between nickelate and STO for all STO/NSNO samples. The elements H, Ni, Ti, Nd, Sr, O profiles are collected from the negative ions H$^-$, Ni$^-$, TiO$^-$, NdO$^-$, SrO$^-$, O$^{2-}$, respectively.

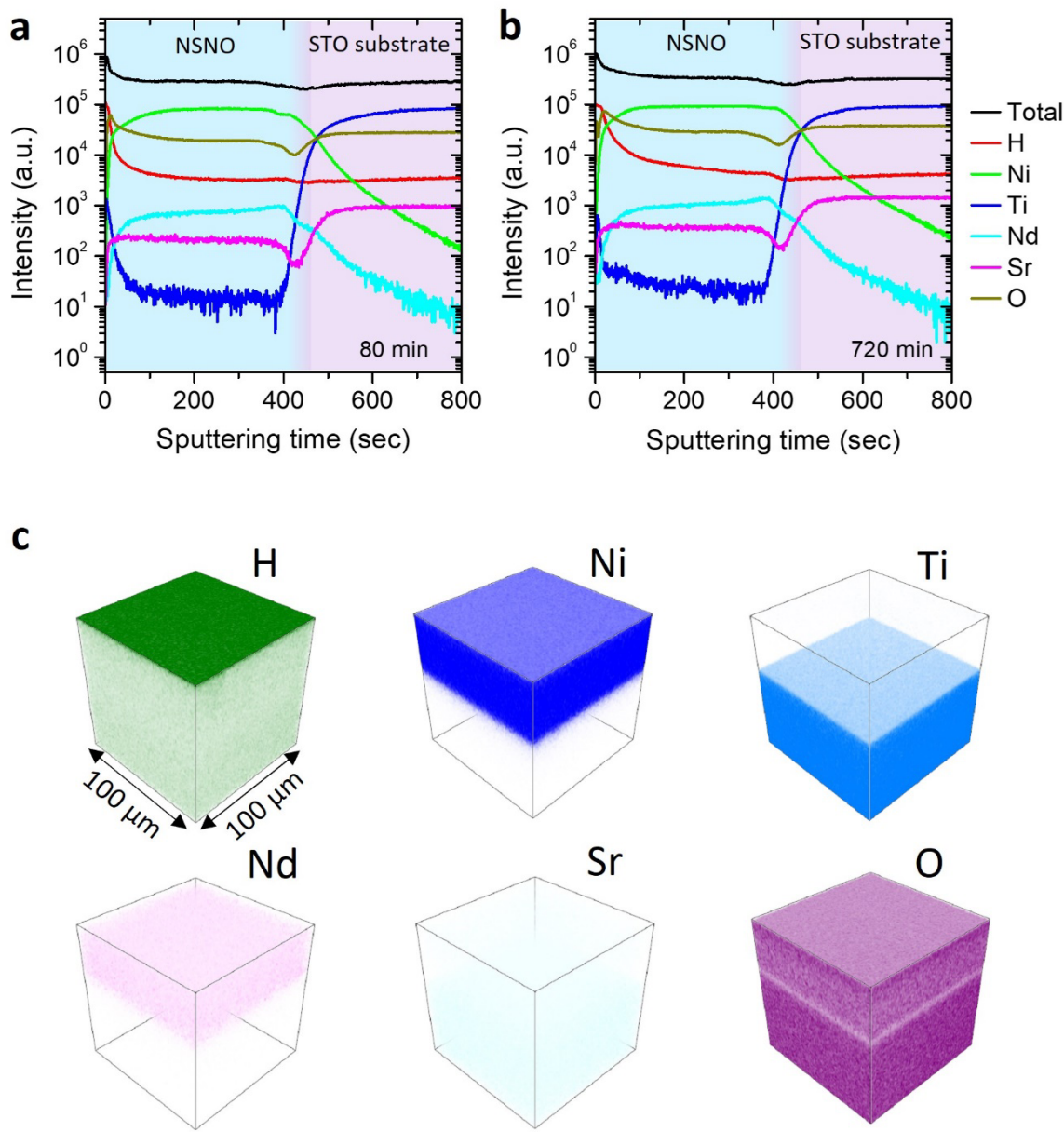

**Supplementary Fig. S3 TOF-SIMS element profiles and 3-dimentional (3D) mapping for 20-nm Nd$_{0.8}$Sr$_{0.2}$NdO$_2$.** SIMS signals of total, H, Ni, Ti, Nd, Sr and O profile for tNSNO at reduction duration of (**a**) 80 min, (**b**) 720 min. (**c**) 3D mappings of H, Ni, Ti, Nd, Sr and O elements for tNSNO sample reduced for 80 min. The element profiles clearly indicate the well-defined interfaces and layer structures between nickelate and STO. The 3D mappings indicate that the elements of the nickelate film are uniformly distributed with no chemical segregation within the measurement size of 100 μm x 100 μm. The elements H, Ni, Ti, Nd, Sr, O profiles are collected from the negative ions H$^-$, Ni$^-$, TiO$^-$, NdO$^-$, SrO$^-$, O$^{2-}$, respectively.

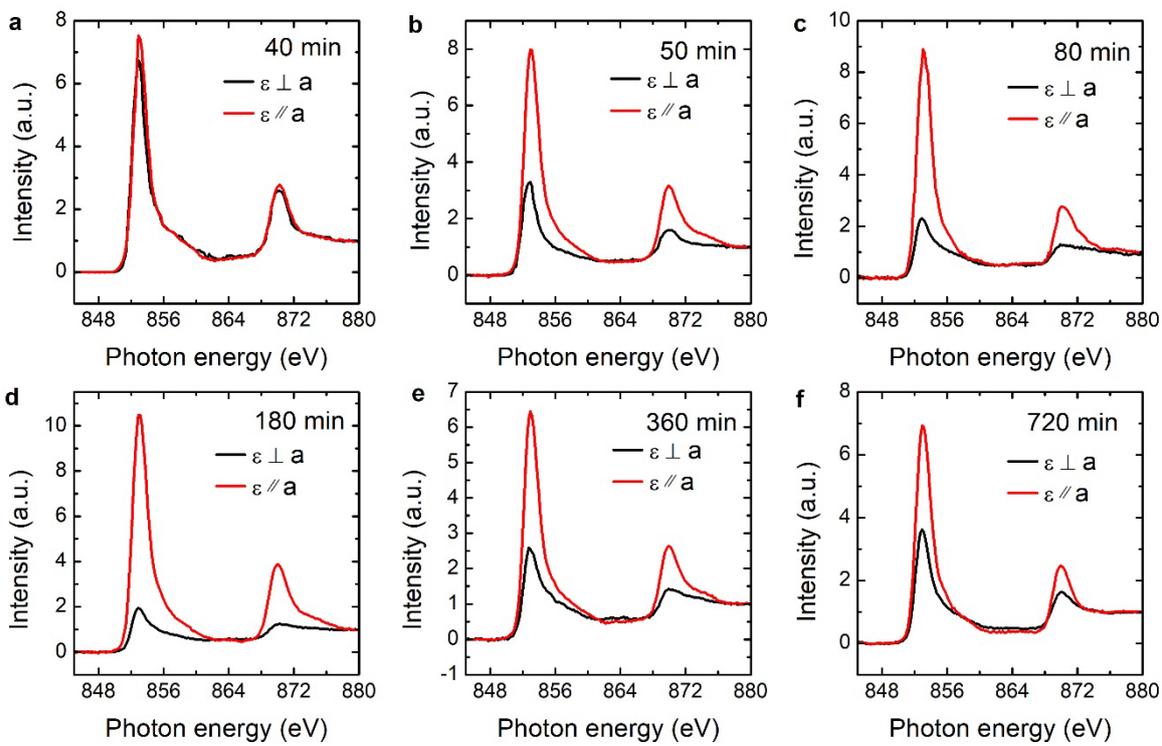

**Supplementary Fig. S4 Reduction dependence of the X-ray absorption spectroscopy.** Ni $L_{2,3}$ edge XAS of STO/NSNO samples for reduction time of (**a**) 40 mins, (**b**) 50 mins, (**c**) 80 mins, (**d**) 180 mins, (**e**) 360 mins and (**f**) 720 mins. The spectra were collected at incident angle of 20° ($\varepsilon \perp a$, light polarization $\varepsilon$ vector is mostly perpendicular to the sample surface) and 90° ($\varepsilon \parallel a$, light polarization $\varepsilon$ vector is parallel to the sample surface).

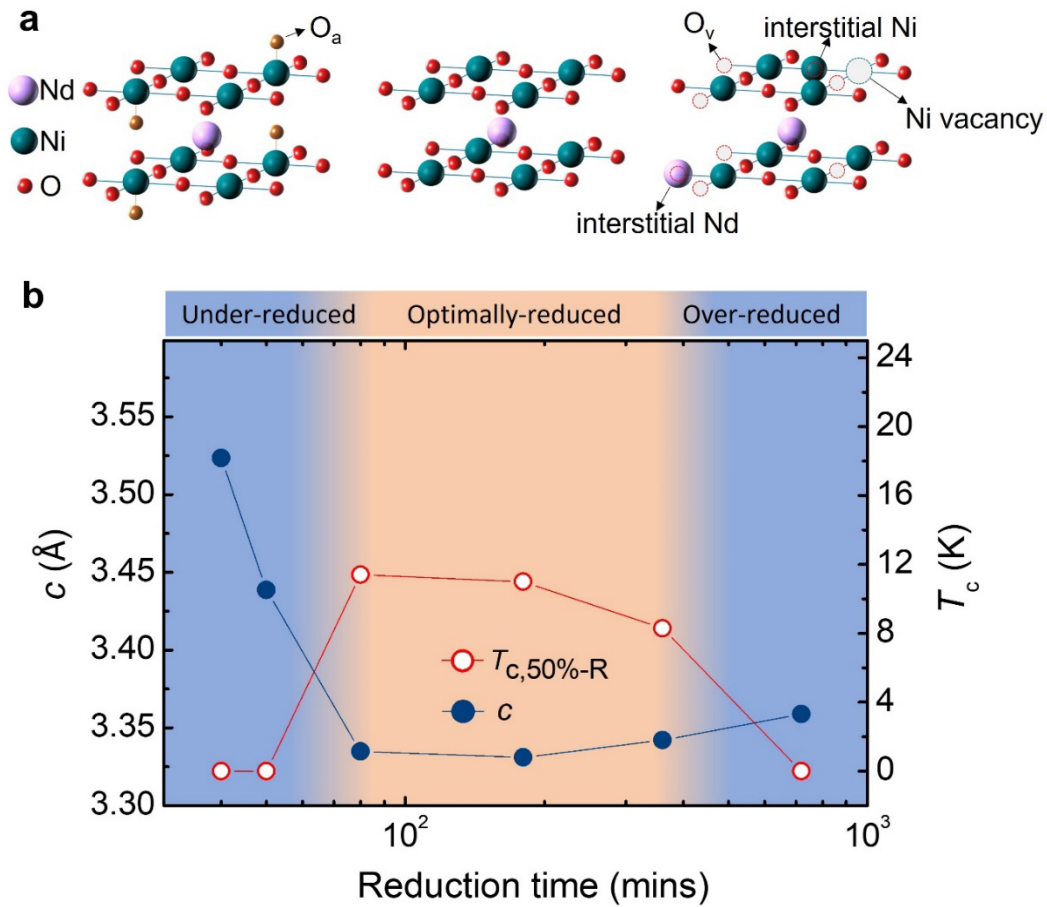

**Supplementary Fig. S5 Schematic of lattice structures and the reduction phase diagram of infinite-layer Nd$_{0.8}$Sr$_{0.2}$NiO$_2$ thin films.** (**a**) Lattice structures in the under-reduced (left), optimally reduced (middle) and over-reduced conditions (right). Perfect NiO$_2$ square planar coordination could be obtained in optimally reduced regime, while residual apical oxygen (O$_a$) and oxygen vacancy (O$_v$) within NiO$_2$ plane are seen in under-reduced and over-reduced regimes, respectively. At the over-reduced condition, formation of oxygen vacancies in the NiO$_2$ plane cause the interstitial Ni/Nd into oxygen vacancy sites. (**b**) The *c*-axis lattice constant, *c*, and the superconducting critical temperature $T_c$ plotted against reduction duration for STO/NSNO samples. The $T_{c\,50\%\text{-}R}$ is defined to be the temperature at which the resistivity drops to 50 % of the value at 20 K (the onset of the superconductivity). The reduction time is in logarithmic scale. The correlation between the lattice constant and superconducting temperature is clearly seen. The *c*-axis lattice constant shows a reversed doom-like dependence with reduction time, as compared to the $T_c$.